\documentstyle[amssymb,prb,aps,twocolumn,psfig,floats]{revtex}

\begin{document}
\title{One dimensional strongly interacting Luttinger liquid of lattice spinless
fermions\thanks{%
Dedicated to Renat Bariev's memory}}
\author{Igor N. Karnaukhov$^{1\dagger }$ and Alexander A. Ovchinnikov$^{1,2}$}
\address{$^{1}$Max-Planck Institut f\"{u}r Physik Komplexer Systeme, N\"{o}thnizer\\
Stra\ss e 38, D-01187 Dresden, Germany\\
$^{2}$ Joint Institute of Chemical Physics, Kosigin St. 4, 117334 Moscow,\\
Russia \\
}
\date{today}
\maketitle

\begin{abstract}
We consider the spinless fermion model with hard core repulsive potential
between particles extended on a few lattice sites $\delta $. The Luttinger
liquid behavior is studied for the different values of a hard core radius.
We derive a critical exponent $\Theta $ of the one particle correlation
function $\langle c_{i}^{\dagger }c_{j}\rangle $ for an arbitrary electron
density and coupling constant. Our results show that at the high density the
behavior of fermions can be described as a strongly interacting Luttinger
liquid with $\Theta >1$. As a result, the residual Fermi surface disappears.
\end{abstract}

\pacs{71.10.Fd; 71.10.Pm}



\smallskip The anisotropic one dimensional (1D) S=$\frac{1}{2}$ Heisenberg
model or the XXZ spin-chain is one of the simple integrable models of the
condensed matter physics \cite{1}. The conception of Luttinger liquid
introduced by Haldane \cite{h} is a very productive and useful for the
description of one- dimensional systems \cite{v} where power-law
singularities in various correlators take place. The correlation lengths of
spin correlation function and one-particle correlation function of spinless
fermions at finite magnetic field and finite temperature have been
calculated by means of a numerical study of the quantum transfer matrix \cite
{2}. Authors have obtained a nonanalytic temperature dependence of
correlation length of the spin correlator $\langle \sigma _{i}^{z}\sigma
_{j}^{z}\rangle $ and the Fermi momentum. They have found in this case a
crossover temperature which depends on coupling constant and filling.
Another scenario of a behavior of the one-particle correlation function has
been obtained in the framework of the integrable one dimensional models with
hard-core repulsive potential \cite{a} . In Refs. \cite{a} we discussed a
strongly interacting Luttinger liquid state at high electron density
characterized by a large value of critical exponent $\Theta $ for the
momentum distribution function at T=0 ( $\Theta >1$ ). In this case the
residual Fermi surface disappears. A critical density {\it n}$_{c}$ ($\Theta
(n_{c})=1)$ separates a Luttinger liquid at a low electron density and an
insulator phase that takes place at an extreme density $n_{\max }=\frac{1}{%
1+\delta }$ . Note, that a strongly interacting Luttinger liquid state and a
high-T$_{c}$ superconducting phase are realized at a small doping.

In this Letter we consider the behavior of spinless fermions in the
framework of the extended version of the XXZ spin-chain proposed by Alcaras
and Bariev recently \cite{aib}. As a concrete example we take the spinless
fermion model with a hard core repulsive interaction. The model Hamiltonian
contains a generalized projector that forbids configurations with two
particles locating at distances less than or equal to $\delta $. In this
case the particles interact only when they are at the closest possible
positions (i,i+$\delta $ +1, $\delta $ is measured in units of the lattice
spacing parameter). The case $\delta =0$ corresponds to the traditional XXZ
spin-chain or the spinless fermion model. As we will see below, at high
electron density the repulsive hard core potential leads to formation of a
strongly interacting Luttinger liquid state without a residual Fermi
surface. This example manifests a new many body state of 1D fermions, named
as a strongly interacting Luttinger liquid \cite{a}. As in Refs. \cite{a},
where this conception has been applied to different integrable models, we
will discuss asymptotic behavior of the one-particle correlation function as
a function of the electron density for several values of the coupling
constant and the core radius. In the low temperature limit the critical
exponents are evaluated of the scaling dimensions from finite size
corrections to the energy spectra and can be calculated using the
thermodynamic Bethe equations. The behavior of XXZ-spin chain depends on the
value of an anisotropic exchange interaction $\Delta $: in the absence of an
external magnetic field the point $\Delta =1$ separates the gap and gapless
antiferromagnetic states. We shall consider both cases, namely $\Delta <1$
and $\Delta >1$, comparing a strongly interacting Luttinger liquid behavior
in the insulator and metal states.

The model can be considered as a generalization of the well-known spinless
fermion model with the Hamiltonian 
\begin{equation}
{\cal H}=\sum_{i=1}^{L}{\cal P}_{\delta }(c_{i+1}^{\dagger
}c_{i}+c_{i}^{\dagger }c_{i+1}+\Delta n_{i}n_{i+1+\delta }){\cal P}_{\delta
},
\end{equation}
where $c_{i}^{\dagger }$ and $c_{i}$ are spinless creation and annihilation
operators of fermions at lattice site {\it i}, the hopping integral equal to
unit and the coupling constant $\Delta $ is dimensionless, ${\cal P}_{\delta
}$ is the projector forbidding two particles at distances less than or equal
to $\delta $. By $n_{i}=c_{i}^{\dagger }c_{i}$ we denote the number operator
for particles on site {\it i}. The system consists of $N$ particles on the
chain with $L$ sites ($L$ is assumed to be even). The Hamiltonian (1) is
transformed into XXZ-chain using Jordan-Wigner representation .

Then, model (1) was exactly solved by the Bethe ansatz method \cite{aib}.
The two-particle scattering matrix is multiplied by an additional scattering
phase shift due to the hard core potential

\begin{equation}
S({k_{i}},{k_{j}})=\exp [-i\delta ({k_{i}}-{k_{j}})]S_{H}({k_{i}},{k_{j}}),
\end{equation}

where $S_{H}({k_{i}},{k_{j}})$ is the two-particle scattering matrix of the
XXZ spin-chain 
\begin{equation}
S_{H}({k_{i}},{k_{j}})={\frac{{1+\exp i(k_{i}+k_{j})+\Delta \exp ik_{j}}}{%
1+\exp i({k_{i}}+{k_{j}})+\Delta \exp i{k_{i}}}}.
\end{equation}

At $\delta =0$ the two-particle scattering matrix (2) is reduced to the one
for the traditional XXZ chain. In the present Letter we consider two
repulsive critical regimes 0$\leqslant \Delta <1$ and $\Delta >1$,
introducing the parametrization for the coupling constant $\Delta =\cos \eta 
$ (0%
\mbox{$<$}%
$\eta \leqslant \frac{\pi }{2}$) and $\Delta =\cosh \eta $ , respectively.

For 0$\leqslant \Delta <1$ the charge rapidities $\lambda _{j}(j=1,...,N)$
related to the momenta of particles ${k_{j}}$ $\left( \exp (i{k_{j})=}\frac{%
\sinh \frac{1}{2}(\lambda _{j}+i\eta )}{\sinh \frac{1}{2}(\lambda _{j}-i\eta
)}\right) $~are obtained by solving the Bethe ansatz equations 
\begin{eqnarray}
\left( \frac{\sinh \frac{1}{2}(\lambda _{j}+i\eta )}{\sinh \frac{1}{2}%
(\lambda _{j}-i\eta )}\right) ^{L-\delta N} &=&(-1)^{N}\exp (-i\delta P) 
\nonumber \\
&&\prod_{j=1}^{N}\frac{\sinh \frac{1}{2}(\lambda _{j}-\lambda _{i}+2i\eta )}{%
\sinh \frac{1}{2}(\lambda _{j}-\lambda _{i}-2i\eta )}{,}
\end{eqnarray}
where $P=\sum_{i=1}^{N}{k_{i}}$ is the momentum.

In terms of the rapidities ${\lambda _{i}}$ the energy of the eigenvalues is
given by 
\begin{equation}
E=2N\cos \eta -2\sin ^{2}\eta \sum_{j=1}^{N}\frac{1}{\cosh \lambda _{j}-\cos
\eta }.
\end{equation}

In the thermodynamic limit the ground state is described by the real
rapidities \{$\lambda _{j}$\}. Taking the thermodynamic limit we obtain the
integral equation of the Fredholm type for the distribution function $\rho
(\lambda )$ for the variable $\lambda $,

\begin{equation}
\rho (\lambda )+\int_{-\Lambda }^{\Lambda }d\lambda ^{\prime
}K_{2}^{<}(\lambda -\lambda ^{\prime })\rho (\lambda ^{\prime })=(1-\delta
n)K_{1}^{<}(\lambda ),
\end{equation}

where the kernel $K_{n}^{<}(\lambda )$ is,

\begin{equation}
K_{\nu }^{<}(\lambda )=\frac{1}{2\pi }\frac{\sin (\nu \eta )}{\cosh \lambda
-\cos (\nu \eta )}.
\end{equation}

The $\lambda $-Fermi level denoted as $\Lambda $ controls the band filling
and the density of fermions is defined by

\begin{equation}
n=\int_{-\Lambda }^{\Lambda }d\lambda \;\rho (\lambda ),
\end{equation}

for $0\leqslant n\leqslant n_{0}$ and

\begin{equation}
1-(1+\delta )n=\int_{-\Lambda }^{\Lambda }d\lambda \;\rho (\lambda ),
\end{equation}

for $n_{0}<n\leqslant n_{\max }$ . The particle density $n_{0}$ corresponds
to a 'half-filling' $\frac{1}{2+\delta }$ and $n_{\max }$ is an extreme
density that corresponds to the full band.

We remind the known results of the one-particle correlation function $%
\langle c^{\dagger }(x)c(0)\rangle $ at T=0 obtained in the framework of
conformal field theory. The correlation function shows an oscillatory
behavior and power-like decay with a scaling dimension $\Delta ^{\prime }$

\begin{equation}
\langle c^{\dagger }(x)c(0)\rangle \simeq \cos (k_{F}x)x^{-2\Delta ^{\prime
}},
\end{equation}

here $k_{F}$ is the Fermi momentum. The momentum distribution function close
to $k_{F}$ is determined by the exponent $\Theta $, 
\begin{equation}
\langle n_{k}\rangle \simeq \langle n_{k_{F}}\rangle -const|k-k_{F}|^{\Theta
}sgn(k-k_{F}),
\end{equation}
where $\Theta =2\Delta ^{\prime }-1$ and 
\begin{equation}
\Theta ={\frac{{1}}{\alpha }}\left( 1-{\frac{{\alpha }}{2}}\right) ^{2}.
\end{equation}

The long-distance power-law behavior of the spin correlator $\langle \sigma
_{i}^{z}\sigma _{j}^{z}\rangle $ is described by the critical exponent $%
\alpha $; $\alpha $=2$\zeta ^{2}(\Lambda )$ is defined by the dressed charge 
$\zeta (\lambda )$ at the $\lambda $ -Fermi level, $\zeta (\lambda )$ is the
solution of the following integral equation 
\begin{equation}
\zeta (\lambda )+\int_{-\Lambda }^{\Lambda }d\lambda ^{\prime
}K_{2}^{<}(\lambda -\lambda ^{\prime })\zeta (\lambda ^{\prime })=(1-\delta
n).
\end{equation}

\begin{figure}[tbph]
\caption{Exponent $\alpha $ as a function of the electron density for $\eta
=0.1$ (solid lines), $\pi /4$ (dashed); $\pi /3$ (dashed with points) and $%
\delta =0,1,2,3$. Result for $\delta =0$ is plotted (dotted line) for
comparision. }
\label{Fig1}
\end{figure}

\begin{figure}[tbph]
\caption{The exponent $\Theta $ , as in Fig. 1.The line separates the
strongly interacting Luttinger liquid state. }
\label{fig2}
\end{figure}

The critical exponent $\alpha $ is analytically calculated for the densities 
$n=~0$ ($\alpha =2$ ), $n_{0}$ $\left( \alpha =4\frac{\pi }{\pi -\eta }%
n_{0}^{2}\right) $ and $n_{\max }$ $\left( \alpha =2n_{\max }^{2}\right) $.
This was done using Wiener-Hopf method. Solving numerically the integral
equations (6), (13) and taking into account conditions (8),(9) we show on
Figs1 and 2 the critical exponents $\alpha $ and $\Theta $ as a function of
the density for some values of $\Delta $ $(\eta =0,1;\pi /4;\pi /3)$ and $%
\delta =0,1,2,3$ . The results of calculations obtained for the traditional
XXZ chain ($\delta =0$ ) are plotted also for comparison in figures using
dotted lines. In low density limit $n\rightarrow 0$ $\alpha $ converges to
the value 2, which is the same as for non-interacting fermions. In the high
density limit $n$ $\rightarrow n_{\max }$ the value of $\alpha $ is
independent on the coupling constant $\alpha =2n_{\max }^{2}$ . In this case
fermions are frozen and can be described by the second term in (1) .  The
model Hamiltonian does not depend on the coupling constant (so, we can
choose it equals unit) and, as a result, the values of the critical
exponents $\alpha $ and $\Theta $ are independent on $\Delta $ (see Figs 3
and 4 also). For $\delta >0$ a density of holes changes in the interval from 
$\delta $/(1+$\delta )$ to 1 whereas the density of fermions varies from
1/(1+$\delta $ ) to 0. The hard core potential makes essential changes in
the structure of the hole state. Note that this interaction also destroys
the hole-particle symmetry.

For $\delta >0$ the function is non-symmetric with respect to 'half-filling'
point and the value of $\alpha $ is less than 1 at $n>0.5$. The correlation
effects obtained due to the hard-core potential were most impressively
displayed in the $\Theta $ behavior. According to (12) small values of the
correlation exponent $\alpha $ leads to a large value of the $\Theta $
correlation exponent (see Fig.2). Comparing the behavior of $\Theta $ at $%
\delta =0$ and $\delta \gneqq 0$ we observe an extremely large value of $%
\Theta $ in the high density region which reaches $\frac{1}{2}(n_{\max
}^{-1}-n_{\max })^{2}$ at maximal possible density $n_{\max }.$ Instead of
the trivial value of $\Theta =0$ that takes place at $\delta =0$ we obtain $%
\frac{1}{2}\left( \frac{3}{2}\right) ^{2}$ at $\delta =1$ , $\frac{1}{2}%
\left( \frac{8}{3}\right) ^{2}$ at $\delta =2$ and $\frac{1}{2}\left( \frac{%
15}{4}\right) ^{2}$ at $\delta =3$ . The value of $\Theta $ increases with $%
\Delta $ and has a maximum value at a 'half-filling' for a large value of
the coupling constant $\Delta $. In Fig.2 we separate by a dotted line the
region of density that corresponds to a strongly interacting Luttinger
liquid state with $\Theta >1$. The critical exponent of the one-particle
correlation function (10) also has a large value in this region ( $2\Delta
^{\prime }>2$).

Using an analogous parametrization for the momenta $\exp (i{k_{j})=}\frac{%
\sin \frac{1}{2}(\lambda _{j}+i\eta )}{\sin \frac{1}{2}(\lambda _{j}-i\eta )}
$ we obtained the Bethe equations for another region of the interaction
coupling $\Delta =\cosh \eta >1$ in the form

\begin{eqnarray}
\left( \frac{\sin \frac{1}{2}(\lambda _{j}+i\eta )}{\sin \frac{1}{2}(\lambda
_{j}-i\eta )}\right) ^{L-\delta N} &=&(-1)^{N}\exp (-i\delta P)  \nonumber \\
&&\prod_{j=1}^{N}\frac{\sin \frac{1}{2}(\lambda _{j}-\lambda _{i}+2i\eta )}{%
\sin \frac{1}{2}(\lambda _{j}-\lambda _{i}-2i\eta )}{.}
\end{eqnarray}

For the distribution function $\rho (\lambda )$ and the dressed charge $%
\zeta (\lambda )$ we have universal integral equations (6), (13) with a new
kernel

\begin{equation}
K_{\nu }^{>}(\lambda )=\frac{1}{2\pi }\frac{\sinh (\nu \eta )}{\cosh (\nu
\eta )-\cos (\lambda )}.
\end{equation}

\begin{figure}[tbph]
\caption{Exponent $\alpha $ as a function of the electron density for $\eta
=0.1$ (solid lines), 0.5 (dashed); 1 (dashed with points) and $\delta
=0,1,2,3$. Result for $\delta =0$ is plotted (dotted line) for comparision. }
\label{fig1}
\end{figure}

\begin{figure}[tbph]
\caption{The exponent $\Theta $ , as in Fig. 3.The line separates the
strongly interacting Luttinger liquid state. }
\label{fig4}
\end{figure}

In Figs 3,4 we show results of calculations of the correlation exponents as
a function of the electron density for different coupling constants $\Delta $
and $\delta $. The behavior of $\alpha $ is similar to the above case. Note
that a minimal value of $\alpha $ is realized for an arbitrary coupling at a
'half-density'. A 'half filling' is a special point in one dimensional
chains. As we note below, the gapless ground state for $\Delta <1$
transforms to the gap antiferromagnetic state for $\Delta >1$ . In Fig 4 the 
$\Theta $ - cusps characterize a gap phase for $\Delta >1$ , a residual peak
takes place for $\Delta <1$ at small $\eta $ only (see a solid line in Fig
2). This leads to the maximum of the value of $\Theta $ for arbitrary $%
\Delta $ and $\delta $ (see the curves obtained for $\eta =0.1;0.5;1$ and $%
\delta =0,1,2,3$ in Fig.4). $\Theta $ is an increasing function of the
parameters of the interactions $\Delta $ and $\delta $. According to
numerical calculation the critical density {\it n}$_{c}$ ($\Theta (n_{c})=1$
) is less then {\it n}$_{0}$. The value of $n_{c}$ depends on the coupling
constant $\Delta $ and the hard-core radius.

A 'half-filling' point is a singular one in the behavior of the critical
exponents. When $\Delta >1$  antiferromagnetic state with a gap takes place
the $\Theta $ has a sharp maximum in this point. The width of the maximum is
defined by a new scale, the gap $\Delta _{c}=4\sinh \eta \frac{K}{\pi }%
k^{\prime }$, where $\eta =\pi K^{\prime }/K$, $k$ and $k^{\prime }$ are
modulus and complementary modulus of Jacobian elliptic functions and
integrals, $K\equiv K(k)$ and $K^{\prime }\equiv K(k^{\prime })$ are
complete and associated complete elliptic integrals of the first kind,
respectively. The $\Delta _{c}$ trends to zero in the $\eta \rightarrow 0$
limit. The elliptic integrals {\it K }and {\it K'} are connected with
coupling constant $\eta $ and define a value of the gap. The gap in this
state is a result of a large magnetic anisotropy. Thus, we can conclude that
a strongly interacting Luttinger liquid is a state taking place near
insulator or bad metal phases. It is possible also that a superconducting
state has some features of a strongly interacting Luttinger liquid.

In summary, using an exact solution of generalized XXZ-spin model we
extended the calculations of a strongly interacting Luttinger liquid state
to the lattice spinless fermion system with a hard-core repulsive potential.
The system has two qualitatively different regions. In the low-density
regime where the hard core potential is not essential we obtain a
traditional Luttinger liquid . In opposite high density limit when the
hard-core potential dominates the state is a strongly interacting Luttinger
liquid with a large value of the critical exponent $\Theta >1$. This
behavior has been illustrated by the numerical calculations. It is shown,
that the value of the critical density that separates these regions depends
on both $\delta $ and $\Delta $. A main feature of this state is an absence
of the residual Fermi surface.

I.K. wishes to thank the support of the Visitor Program of the
Max-Planck-Institut f\"{u}r Physik Komplexer Systeme, Dresden, Germany.

$^{\dagger }$ Permanent address: Institute of Metal Physics, Vernadsky
Street 36, 03142 Kiev, Ukraine{\bf .}

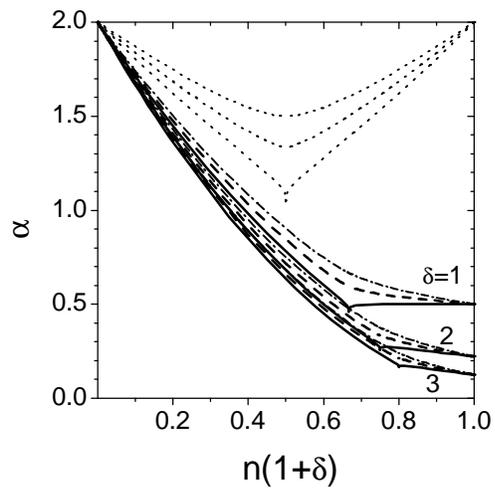
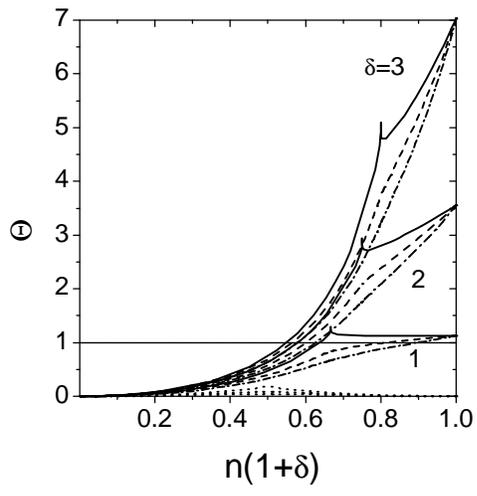
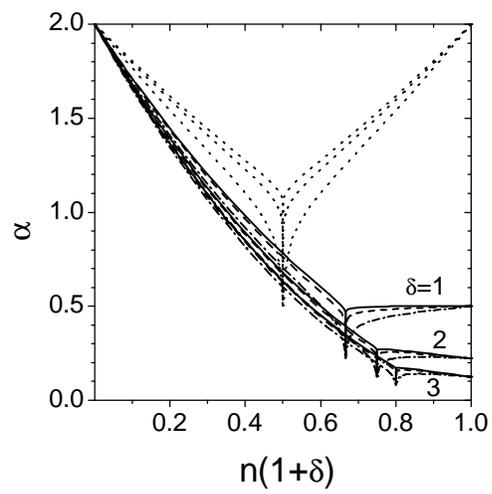
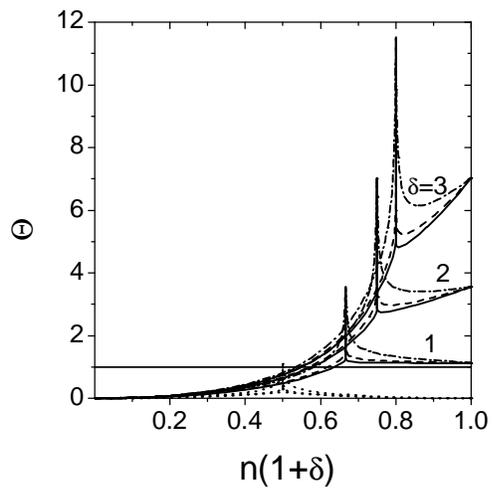
\end{document}